\documentclass{osa-article}

\journal{osajournal}

\usepackage[version=4]{mhchem}
\usepackage{acronym}  
\usepackage{graphicx}%
\usepackage{float}
\usepackage{soul}
\usepackage{subcaption}
\usepackage{textcomp}
\usepackage{makecell}

\usepackage{tabularx}
\usepackage{comment}
\usepackage{siunitx}
\begin{document}

\title{Characterization of noise regimes in Mid-IR free-space optical communication based on quantum cascade lasers}




\author{Marco Seminara\authormark{1,2,$\dagger$}, Tecla Gabbrielli\authormark{1,2,$\dagger$}, Nicola Corrias\authormark{1,2,3}, Simone Borri\authormark{2,1}, Luigi Consolino\authormark{2,1}, Marco Meucci\authormark{2,4}, Paolo De Natale\authormark{2,1}, Francesco Cappelli\authormark{2,1}, and Jacopo Catani\authormark{2,1,*}
}

\address{\authormark{1} European Laboratory for Non-Linear Spectroscopy (LENS), Sesto Fiorentino (FI), Italy \newline 
\authormark{2} National Institute of Optics-CNR (CNR-INO), Sesto Fiorentino (FI), Italy \newline 
\authormark{3}QTI s.r.l., Largo E. Fermi 6, 50125 Firenze, Italy \newline
\authormark{4} ARTES4.0 Competence Center on Robotics and Digital Technologies, Node of Sesto Fiorentino (FI), Italy \newline 
\authormark{$\dagger$} these authors contributed equally to this work \newline 
\authormark{*} Corresponding author: jacopo.catani@ino.cnr.it 
}



\begin{abstract}
The recent development of Quantum Cascade Lasers (QCLs) represents one of the biggest opportunities for the deployment of a new class of Free Space Optical (FSO) communication systems working in the mid-infrared (Mid-IR) wavelength range. As compared to more common FSO systems exploiting the telecom range, the larger wavelength employed in Mid-IR systems delivers exceptional benefits in case of adverse atmospheric conditions, 
as the reduced scattering rate strongly suppresses detrimental effects on the FSO link length given by the presence of rain, dust, fog and haze.
In this work, we use a novel FSO testbed operating at \SI{4.7}{\micro m}, to provide a detailed experimental analysis of noise regimes that could occur in realistic FSO Mid-IR systems based on QCLs. Our analysis reveals the existence of two distinct noise regions, corresponding to different realistic channel attenuation conditions, which are precisely controlled in our setup. To relate our results with real outdoor configurations, we combine experimental data with predictions of an atmospheric channel loss model, finding that error-free communication could be attained for effective distances up to 8~km in low visibility conditions of 1 km.
Our analysis of noise regimes may have a key relevance for the development of novel, long-range FSO communication systems based on Mid-IR QCL sources.
%
\end{abstract}




\section{Introduction}\label{sec:intro}
Free-Space Optical (FSO) links represent a valuable option when the implementation of fiber links is impractical and realizing point-to-point or satellite-assisted communication infrastructures are much more efficient and convenient~\cite{Willebrand:2001}. The technological research on Free-Space Optical Communication Systems (FSOCSs) and reinforcement of the existing infrastructures pave the way not only to the possible replacement of fiber cables in the rising 5G networks~\cite{esmail:2019,Mirabissi}, but also to the development of new technology for the upcoming 6G era, where the implementation of a hybrid FSO/microwave platform can open new horizons for telecommunications~\cite{Klaus:2018,Dang:2020}. Furthermore, indoor optical wireless communication can benefit from the improvement of laser-based FSO technology exploiting the advantages of a higher frequency of the carrier, a wider bandwidth, a much higher spatial directionality, unlicensed operation, high security compared to radio frequencies together with lower costs, and simpler infrastructure with respect to fiber links~\cite{Dehghani:2021,Khan:2017}. Commonly, FSOCSs have been tested and developed in the Near Infrared (NIR) wavelength range (\SIrange{0.75}{3}{\micro \meter}), and in particular in the so-called 
telecom wavelength sub-range (\SI{1.55}{\micro \meter})~\cite{Bloom:2003}, on which the worldwide fiber-based communication infrastructure is currently set. The NIR spectral region is equipped with well-established technologies on both transmitter and receiver sides (e.g., around \SI{800}{\nano \meter}, with silicon detectors or high-power sources such as VCSEL). In the last decades, another spectral region has started to be attractive in terms of FSO links, the Mid Infrared range (Mid-IR, $\lambda> \SI{3}{\micro \meter}$) ~\cite{Bloom:2003}, as Mid-IR atmospheric transparency windows can usefully complement the NIR ones. One of the most attractive features of the Mid-IR is its reduced sensitivity to particle scattering, scintillation, and background noise due to the black-body emission of the Sun (peaked at $\lambda \sim \SI{500}{nm}$ and well suppressed above \SI{3}{\micro \meter})~\cite{faist:1994,Su:18}. Moreover, the high transparency windows around \SI{4.0}{\micro \meter} goes along with a strongly reduced black-body emission of Earth, which is peaked at $\lambda \sim \SI{10}{\micro \meter}$ and is well suppressed for $\lambda < \SI{5}{\micro \meter}$~\cite{faist:1994}. In the Mid-IR, it is also possible to achieve larger transmission efficiency than in the NIR in case of adverse weather conditions (fog, haze, clouds)~\cite{Corrigan:2009, Su:18}, which is relevant also for satellite communications \cite{Flannigan_2022}. 
In this scenario, the advent of Quantum Cascade Lasers (QCLs) with highly-tailorable emission covering the \SIrange[]{3}{12}{\micro \meter} range~\cite{faist:1994,Tombez:2013a,faist:2013,Riedi:2015,cathabard2010quantum}, represented a technological breakthrough for extensive development of Mid-IR FSOCSs. 
Since their invention, the attention of the communication community has been attracted by the very short lifetime (< \SI{1}{ps}) of their lasing transitions, which allow both electrical and optical modulation of the emitted radiation at high frequencies (up to several GHz)~\cite{paiella:2001,hinkov:2016}. Typically used as spectroscopy sources~\cite{Consolino:2018a,borri:2019}, mid-IR QCLs started to be tested also as transmitters in FSO communications~\cite{Corrigan:2009,gutowska:2011,mikolajczyk:2014}. Besides initial proof-of-concept FSOCSs embedding QCLs emitting around \SI{4.7}{\micro \meter} have been reported for distances of about \SI{2}{\meter}~\cite{Liu:2015,Corrias:2022}, recent years saw a massive development of directly-modulated QCL FSOCS working in such favorable wavelength range \cite{spitz2022,pang:2021}. Such effort recently culminated in the capability to attain multi-Gbps bitrates with room-temperature QCLs \cite{pang2022_6gbps}, and the 10-Gbps threshold has recently been overcome by employing 9~$\mu$m QCL sources \cite{Dely:2022}. 
Indeed, the effective deployment of reliable Mid-IR FSOCSs based on QCLs in realistic environments requires that the various noise contributions, which depend on the specific application, are analyzed and evaluated. In this sense, a theoretical model and simulations to study the transmission rate under various atmospheric conditions have been recently proposed\cite{sauvage_outdoor_MIR_link_noise}, considering two different laser sources (\SI{1.55}{\micro \meter} and \SI{4.0}{\micro \meter}) and a fixed distance of \SI{4}{\kilo \meter}. Nonetheless, a thorough experimental characterization of the impact of different noise conditions on the communication performances of a Mid-IR FSO link is still lacking.
To tackle this issue, in this work we exploit a novel testbed, based on a QCL emitting at \SI{4.72}{\micro \meter} to characterize, for the first time, the occurrence of two distinct noise regimes, corresponding to different, realistic conditions of channel attenuation. We analyze the performance of the QCL-based Mid-IR FSOCS in such regimes, highlighting very different behavior for the communication quality as a function of several experimental parameters. Thanks to the tunability of the presented setup, we also explore an intermediate noise regime, observing a clear transition in the Packet Error Rate (PER) trend as the two noise regions are spanned. 
By combining our findings with the predictions of an atmospheric propagation model, we also estimate reliable Mid-IR FSO communications for our system covering distances up to 8~km in scarce visibility conditions (1 km).
%
%
%
\section{Setup overview}
\label{sec:Setup}
\begin{figure}[ht!]
\centering
\includegraphics[width=0.8\columnwidth]{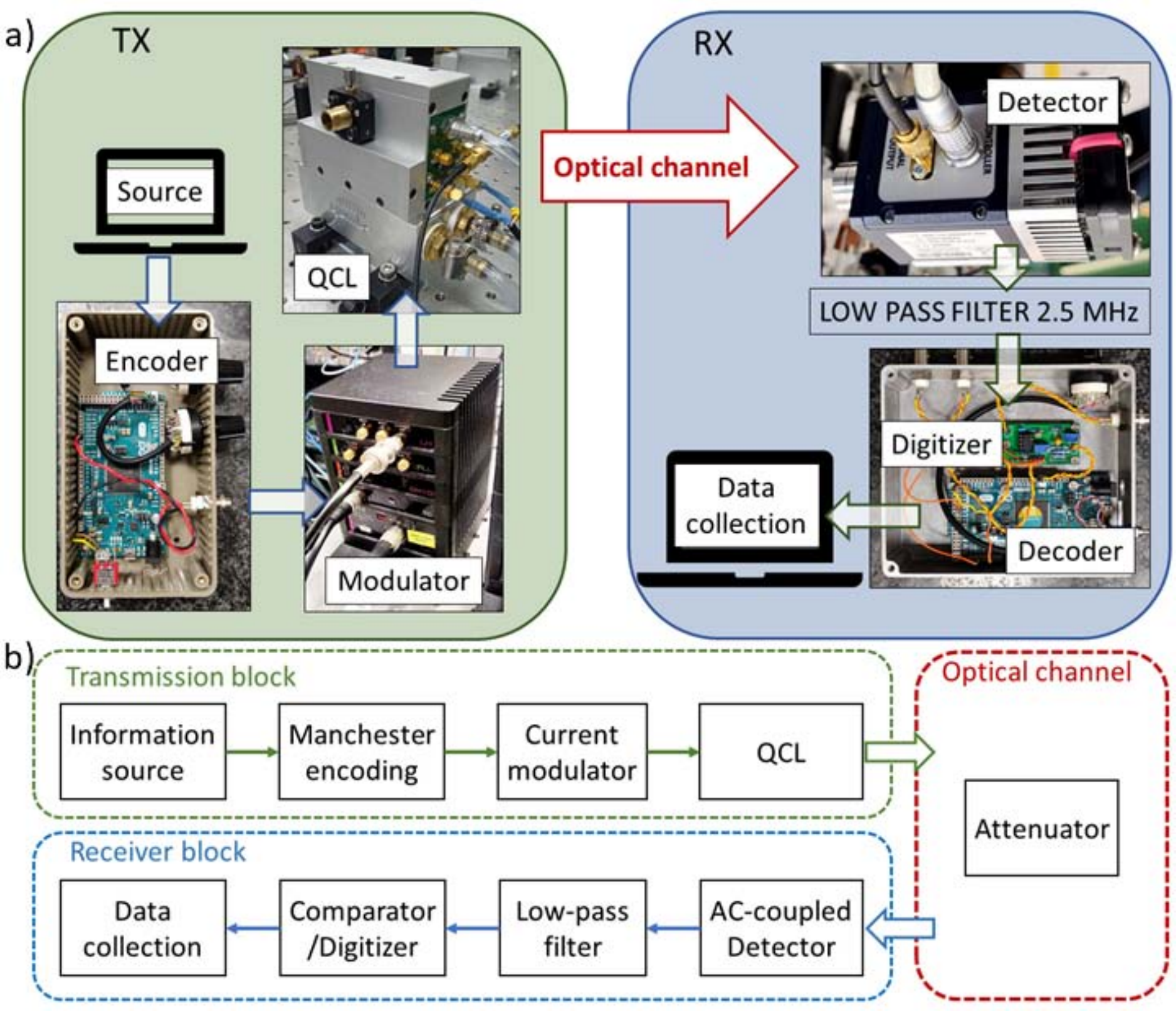}
\caption{Sketch (a) and block diagram (b) of the experimental setup divided into transmission block (green), optical channel (red) and receiver block (blue).
}
\label{fig:1.setup}
\end{figure}

The Mid-IR communication system we employ for the characterization of the noise regimes is composed of two distinct units: the transmitter unit (TX), where a digital message is encoded in the light emitted by the Mid-IR source through amplitude modulation (AM) via a current modulation provided by the current driver, and the receiver stage (RX), where the optical signal is detected, converted to voltage, and digitally processed for message decoding (see Fig.~\ref{fig:1.setup}a).
 
The modulated Mid-IR light propagates in free space passing through a variable optical attenuation system that simulates long-distance channel losses similar to what is reported in \cite{spitz2022}. An AC-coupled amplified detector collects the light at the receiver side. The amplified analog signal is first digitized by a variable-threshold comparator stage. Then, it is decoded and analyzed by the digital RX platform, performing a real-time comparison with a pre-stored reference message. In the following sections, details on both the TX-RX stages and the experimental setup are given.
\subsection{Optical signal generation, controlled attenuation and detection}
The FSOCS source is a custom Fabry-Pérot continuous-wave QCL fabricated at ETH (Swiss Federal Institute of Technology Zurich, Switzerland), with an emitting wavelength of $ \lambda = \SI{4.72}{\micro \meter}$, and working at room temperature ($T=$\SI{18}{\celsius}). The laser works in single-mode regime from threshold current (\SI{638}{mA}) up to \SI{680}{mA}, while for $I>\SI{680}{mA}$ it operates in multi-mode regime. The maximum output optical power achievable in single-mode regime is \SI{21}{mW}, with a driving current of~\SI{680}{mA}. The laser is powered by an ultra-low-noise current driver (QubeCL15-P from ppqSense srl), characterized by a nominal current noise density of $\SI{200}{pA/\sqrt{\mathrm{Hz}}}$. The current driver is equipped with a low-noise current modulator characterized by a maximum modulation amplitude of \SI{5}{mA} and a modulation bandwidth of \SI{0.3}{MHz}. The beam propagates indoor in free-space travelling an optical path length of \SI{3}{ \meter} until it reaches the RX. 
The beam passes through a variable optical attenuator for simulating different attenuation regimes that can occur in a long-distance outdoor FSO communication (Secs. \ref{sec:noise} and \ref{sec:theory}).
By exploiting the linear polarization of the laser light, a variable attenuation is achieved via precise adjustment of a rotating polarizer plate (WP25H-Z holographic wire grid polarizer by Thorlabs). The attenuation used in the characterization covers the range from \SI{14}{dB} to \SI{52}{dB} (Sec.~\ref{sec:result&di})~\cite{note1}. After free-space propagation, the beam is focused on a two-stage transimpedance preamplified Mid-IR \ce{HgCdTe} photovoltaic detector (detector PVI-4TE-5-2x2, preamplifier MIP-10-250M-F-M4 both from Vigo System). The detector has a nominal bandwidth of \SI{180}{MHz} and operates in the wavelength range from \SI{2.5}{\micro \meter } to \SI{5}{\micro \meter}. At \SI{4.72}{\micro \meter }, the detector saturation occurs for an incident power of $\SI{1.2}{m W}$ and its measured quantum efficiency is \SI{33}{\%}~\cite{Gabbrielli:21}. The detector is used in the linear responsivity regime (for $P<\SI{1.2}{mW}$) where the output current is directly proportional to the incident flux of photons. The sensitivity limit of the detection system is given by the detector dark current.
\subsection{Implementation of the digital communication signal}\label{Digital communication signal implementation}
The used hardware is represented in Fig.~\ref{fig:1.setup}a) and it is composed of a TX unit (green block) and an RX unit (blue block). The digital data stream is generated by a digital open-source low-cost microcontroller board (Arduino DUE, the \textit{Encoder} in Fig.~\ref{fig:1.setup}a)). An On-Off Keying (OOK) scheme with Manchester encoding~\cite{Rajagopal:12} is used, which guarantees a constant-average signal. The system can transmit a continuous data stream of 62500 packets with a baud rate up to \SI{115}{kbaud}. We note that larger baud rates could easily be attained by direct modulation of the QCL chip, which was not feasible in the present configuration. However, this is not limiting the breadth of the results on noise characterization.
The packets are composed of 9 bytes, divided into 3 initial equalization bytes for signal pre-equalization, 2 synchronization bytes, and 4 data payload bytes. The digital information is encoded in the beam as intensity modulation via the current driver which adds AC modulation on top of the laser DC driving current. On the receiver side (blue block in Fig.~\ref{fig:1.setup}b)), the signal at the detector output passes through a 2.5 MHz Low-Pass filter (BLP-2.5+ from Mini-Circuits), used to cut-off frequency components higher than 10 times the first harmonic of the modulation signal. The resulting analog signal is then digitized by a variable threshold comparator and, finally, decoded in real-time by a second Arduino DUE board, which compares it with a pre-stored message.
The received signal is recorded via a 2.5 Gs/s 4-channel digital oscilloscope (Tektronix MDO3024 200 MHz). The performance of the system is evaluated in terms of PER, a relevant metric to assess the quality of data-structured digital transmission channels~\cite{Seminara2020,ASHOK:2019, Malik:15, 2019iv, Meucci21}. The PER is calculated as the ratio between the number of received packets with at least one wrong bit and the total amount of sent packets.
We send 62500 packets, chosen as the best compromise between a reasonable measurement time (order of minutes) and an acceptable target PER threshold for error-free communication, corresponding to PER = 1.6 $\times 10^{-5}$. Indeed, assuming a uniform distribution of the erroneous bits on the received packets, the PER can be directly related to the Bit Error Rate (BER) \cite{Khalili05}, as shown in Sec.~\ref{sec:theory}. The resulting BER threshold, $\sim$ 3 $\times 10^{-7}$,  is much smaller than the one required for, e.g., a reliable internet connection after implementation of Forward Error Correction (FEC) codes~\cite{wilson:96}.

\section{Overview of Attenuation and Noise Regimes}
\label{sec:noise}
During the design of FSOCS, it is important to correctly evaluate the link budget and to determine the noise contributions influencing the SNR (Signal-to-Noise Ratio)~\cite{book:kaushal2017}. In addition to the dynamical effects of noise related to the optical signal propagation (e.g. flaring, scintillation, turbulence~\cite{ITU1, ITU2}), the noise in a FSOCS is given by a combination of the intensity noise of the source and the detector background noise. Depending on the optical signal attenuation and on the type of communication (long or short-range, high or low visibility), the FSOCS can operate in different noise scenarios. In this work, we aim at implementing two configurations of communication, corresponding to two realistic attenuation regimes: the High Attenuation Regime (HAR), dominated by propagation and attenuation losses, and the Low Attenuation Regime (LAR), where the intensity noise floor of the laser source prevails on the background noise of detector. As shown in the following sections, for each regime a thorough noise analysis is carried out and the performance of the FSOCS are experimentally studied. 
\begin{figure}[ht!]
\centering
\includegraphics[width= \linewidth]{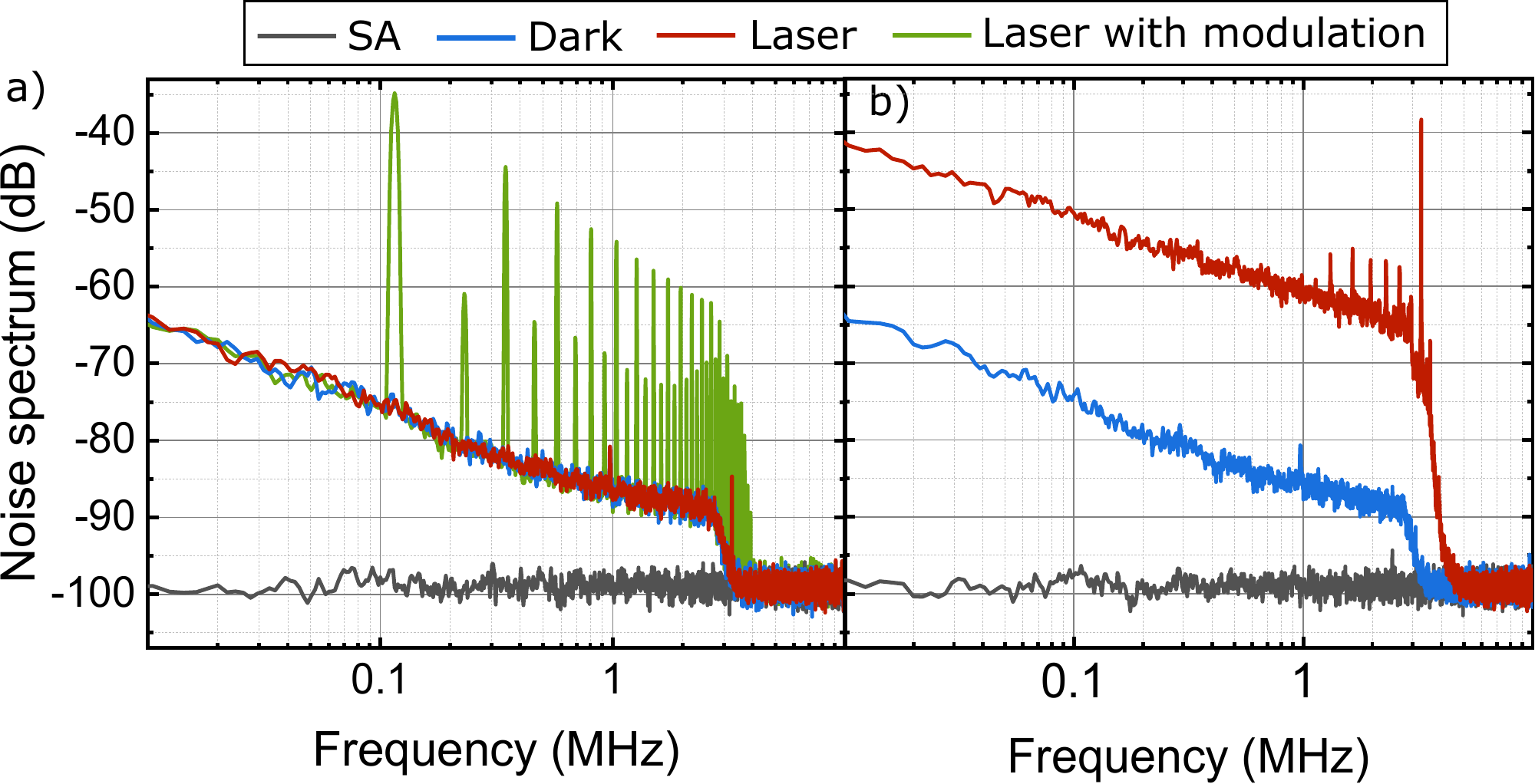}
\caption{Signal noise spectra in the HAR (a) and in the LAR (b). We show the recorded spectrum analyzer background (grey trace), the detector background (blue trace), the laser+detector without modulation (red trace) and applying a digital modulation at \SI{115.2}{kHz} with a MD of 100\% (green trace in the panel a)).
In both graphs the frequency cutoff due to the \SI{2.5}{MHz} low-pass filter is evident.
}
\label{fig:fig2}
\end{figure}
\subsection{Detector-limited noise floor in HAR}
As a first step, the quality of the transmission channel has been assessed in HAR conditions. That may be the case for a long-range outdoor FSO communication where the high attenuation due to absorption and/or scattering by molecules and aerosol particles 
may lead to a very large extinction of the propagating optical beam~\cite{ricklin:2006,henniger:2010}. In this regime, the residual optical power impinging could be as low as few nW's, 
and we expect the main noise contribution to be given by the detector background noise, which can even exceed the intensity noise floor of the laser (Fig.~\ref{fig:fig2}a)). 
To implement this regime, we attenuate the optical power incident onto the detector to obtain a noise floor that is limited by the detector background, so that the laser intensity noise lays below the detector background noise (red and blue trace, Fig.~\ref{fig:fig2}a)). In this configuration only the peaks corresponding to the AM signal (green trace) emerge above the noise floor. For this test, we operate the laser nearby the threshold of the lasing process (driving current $I=\SI{642}{mA}$) so that the relative Modulation Depth (MD), calculated as the ratio between the peak-to-peak amplitude and the DC component of the signal, can reach large values, as it is not limited by the maximum absolute current modulation achievable by our modulator (see Sec.~\ref{sec:noise}). This configuration is the most significant configuration from a standard communication point of view, as it corresponds to a pristine OOK amplitude modulation scheme. Using larger laser currents, instead, would maintain the laser in a stable single-mode operation (well above the lasing threshold in both ON and OFF phases) during the whole transmission process. However, this would limit the relative MD value achievable by our system (and hence the SNR).

Even in HAR condition, where the FSOCS intensity noise floor is fully dominated by the detector background, we are able to detect the AC-modulated signal (green trace, Fig.~\ref{fig:fig2}a)) which, after integration on the receiver bandwidth, corresponds to a SNR of \SI{13}{dB} at the Manchester clock rate frequency of \SI{115.2}{kHz} with a MD of 100\%. 
\subsection{Source-limited noise floor in LAR}
As a second step, we also explore the FSOCS application in the LAR (Fig.~\ref{fig:fig2}b)). In this case, the amount of light collected by the RX stage is large enough that the intrinsic laser intensity noise contribution exceeds the background noise floor (red and blue trace, respectively, in Fig.~\ref{fig:fig2}b)) and the background noise level of the RX stage is not expected to significantly affect the transmission quality. To explore this source-limited scenario, we set the laser current  $I=\SI{662}{mA}$ ($P_{\mathrm{out}}=\SI{12.9}{mW}$) and we test the FSOCS for different MDs with a fixed optical attenuation of \SI{13}{dB}. In this regime the laser is operated in a single-mode regime (above threshold), reducing intensity and frequency fluctuations due to small thermal instabilities. This regime could be relevant in FSOCSs with small channel losses, e.g., in good weather conditions, for short-range communication or/and in a controlled environment such as indoor FSO wireless communication.  As shown in Fig.~\ref{fig:fig2} b), in these working conditions the detected noise floor lies well above the background noise (up to \SI{20}{dB}) and it is dominated by the QCL intensity noise~\cite{Zhao:2019}.
The QCL intensity noise spectrum features the typical $1/f$ trend of the flicker noise, characterizing this type of lasers~\cite{borri2011frequency,bartalini2011,Zhao:2019}.
\section{Theoretical overview}
\label{sec:theory}
\subsection{Signal-to-Noise Ratio and Packet Error Rate}
The communication system performance can be characterized in terms of PER and SNR.
For an OOK modulation, despite the noise spectrum of the system shows a global flicker noise shape (see Fig.~\ref{fig:fig2}a), due to the limited bandwidth over which the detection is performed, the noise can be safely approximated by an Additive White Gaussian Noise (AWGN) spectrum around the baseband frequency of \SI{115.2}{KHz}. Under this assumption, the bit-error probability depends on the SNR through the well-known Q-function~\cite{Stern:04}: BER = Q$ \left(\sqrt{\mathrm{SNR}} \right)$.
Assuming a uniform distribution of errors, the PER, in turn is connected to the BER through PER = $1- (1-\mathrm{BER})^\mathrm{N} = 1- \left(1-Q \left (\sqrt{\mathrm{SNR}}\right) \right)^\mathrm{N}$ \cite{Khalili05}, where N is the number of the packet bits. Hence, PER can be related to the SNR. This latter parameter is related to the measured quantities via SNR(dB) = $20~\mathrm{log} \left(S_{\mathrm{RX}}/2\sigma_{\mathrm{RMS}}\right)$, where $S_{\mathrm{RX}}$ is the received peak-to-peak AC signal, and $\sigma_{\mathrm{RMS}}$ represents the Root Mean Square (RMS) of the noise level.
%
%
In order to relate our experimental investigation with realistic FSO communication conditions, where the commonly-used parameter is channel attenuation, we write the Optical Attenuation (OA) of the FSO channel as: 
\begin{equation}
 \label{eq:3}
   \mathrm{OA} (\mathrm{dB}) = -10~\mathrm{log} \left(\frac{P_{\mathrm{inc}}}{P_{\mathrm{out}}}\right) = - 10~\mathrm{log} \left(\frac{S_{\mathrm{RX}}}{G \cdot R\cdot MD \cdot P_{\mathrm{out}}}\right) ,   
\end{equation}
where G = 26.5 is the gain of the AC transimpedance stage of the detector, $R = \SI{2793}{V/W}$ is the responsivity of the detector, and $P_{\mathrm{out}}$ is the optical power emitted by the QCL operating in single-mode. In particular, the factor $S_{\mathrm{RX}}/(G \cdot R\cdot MD)$ is equal to the incident power onto the detector, $P_\mathrm{inc}$. In order to characterize the performance of our FSOCS, we find convenient to define the Maximal Optical Attenuation (MOA) as the largest tolerable channel attenuation to attain a defined threshold PER value. In quantifying the MOA we consider optical power as the maximum yet guaranteeing stable single-mode operation of the QCL ($P_{\mathrm{max}}=\SI{21}{mW}$). It is possible to estimate the MOA by replacing $P_{\mathrm{out}}$ with $P_{\mathrm{max}}$ in Eq. \ref{eq:3}.
In the LAR regime we study the PER for different MD values, each labeled by the index $i$. Fixing $P_{\mathrm{max}}$, $G$ and $R$ gives a constant ratio $S_{RX} / \sigma_{\mathrm{RMS}}=k_i$, since both the terms are proportional to the residual optical power collected by the RX stage after the optical attenuation stage. In the HAR regime we study the PER as the MOA varies. Therefore, it is useful to rewrite the SNR as a function of MOA, considering $P_\mathrm{out}= P_\mathrm{max}$.
This yields the following set of relations:
\begin{equation}\label{eq:combined}
    \mathrm{SNR}_i(\mathrm{dB})=\begin{cases} k_{\mathrm{i}}  & \text{in LAR,} \\D_i(\mathrm{dB}) -2~\mathrm{MOA(dB)}    & \text{in HAR,}\end{cases}
\end{equation}
where $D_\mathrm{i} = 20 \log \left ( \dfrac{G \cdot R \cdot MD \cdot P_{\mathrm{max}} }{2 \sigma_{\mathrm{RMS}}} \right )$ is constant in our measurement conditions, and represents the $SNR_{\mathrm{i}}$ for negligible channel attenuation.
\subsection{Modeling outdoor FSO links}
\label{sec:model}

\begin{figure}[ht]
\centering
\includegraphics[width=0.8\textwidth]{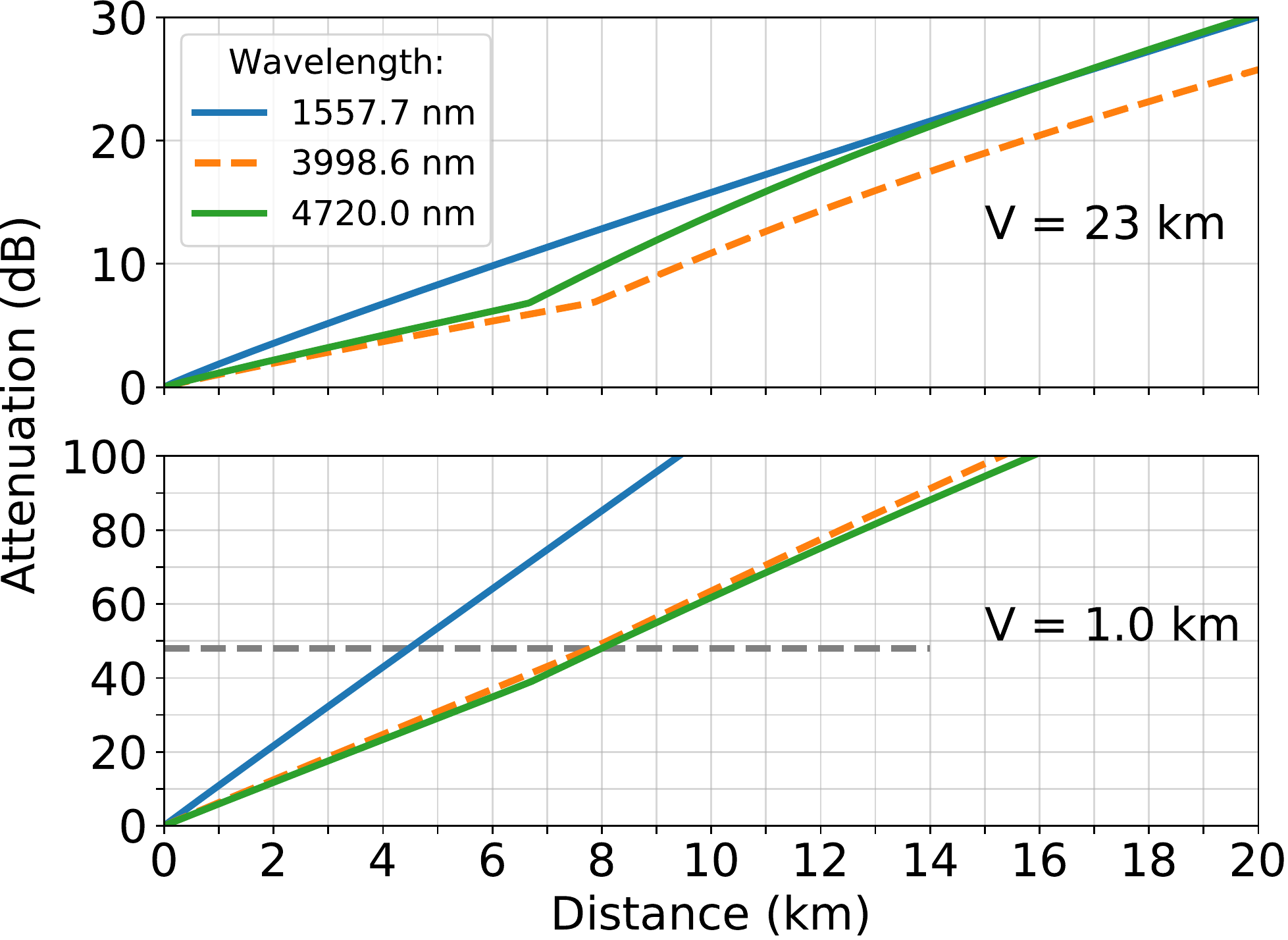}
\caption{Simulation of the attenuation as a function of the communication distance for a terrestrial point-to-point FSO communication at sea level, for very clear-air condition ($V = \SI{23}{\kilo \meter}$ (upper graph), and adverse weather conditions ($V = \SI{1}{\kilo \meter}$ (lower graph). In the graphs, the attenuation due to scattering, atmospheric absorption (HITRAN database \cite{hitrandat,hotw,Rothman:2013}), beam divergence (geometrical attenuation) and scintillation are considered. We assume to have both a receiver and a transmitter with an optical aperture radius of \SI{10}{cm}, and that the geometrical losses start affecting the signal after twice the Rayleigh length.
The dashed gray line represents the error-free communication limit characterizing our system. 
}
\label{fig:fig3}
\end{figure}

In order to relate the retrieved data to realistic outdoor FSO communication scenarios, it is necessary to model and simulate common outdoor conditions in terms of experimentally accessible parameters. In an outdoor FSO link, the propagating beam is attenuated by atmospheric factors such as particle scattering (e.g. by molecules, aerosols, dust, smoke), molecular absorption, and weather conditions (rain, mist, snow, and fog)~\cite{henniger:2010}. In addition, the quality of the received signal can be also affected by geometrical factors such as beam divergence~\cite{henniger:2010}. 
Regarding the atmospheric attenuation, we simulate a simplified scenario that considers particle scattering (i.e. molecules, aerosol), absorption, and scintillation due to turbulence~(Fig.~\ref{fig:fig3}). In these conditions, the atmospheric attenuation coefficient due to scattering and absorption  is described as  ~\cite{book:kaushal2017}:
 $
\gamma(\lambda)=\alpha_{\mathrm{m}}(\lambda)+\alpha_{a}(\lambda)+\beta_{\mathrm{m}}(\lambda)+\beta_{\mathrm{a}}(\lambda),
$
where $\lambda$ is the laser wavelength, $\alpha_{\mathrm{m}}(\lambda)$ and $\alpha_{\mathrm{a}}(\lambda)$ are the molecular (m) and aerosol (a) absorption coefficients, respectively, while $\beta_{\mathrm{m}}(\lambda)$ and $\beta_{\mathrm{a}}(\lambda)$ are the scattering ones. It is difficult to give a precise \textit{a priori} estimation of absorption coefficients, as they depend on the gaseous composition of the air, which can vary consistently with the specific scenario. For instance, the composition varies at different altitudes and/or latitudes, as well as for different seasons and environments (e.g. countryside, city, desert, sea). In our work, we estimate the absorption coefficient by using the atmospheric model named \textit{USA model, mean latitude, summer, $H=0$} of the HITRAN database~\cite{hitrandat,hotw,Rothman:2013}, where $H=0$ is the altitude (sea level).
In the simulation, we consider both Rayleigh and Mie scattering types. The former describes the scattering due to particles with a radius $r << \lambda $ (e.g. molecules). The latter describes the scattering due to aerosol (like fog, clouds and haze) where $r \simeq \lambda$~\cite{book:kaushal2017}. 
We use the formula of the LOWTRAN code for the Rayleigh scattering attenuation due to molecules~\cite{kneizys1980atmospheric}. 
The attenuation coefficient $\beta_{\mathrm{a}}(\lambda)$ due to aerosol is calculated as a function of the visibility $V$ (expressed in km), where $V$ is defined as the distance at which the optical power of a propagating beam of visible green light ($\lambda_0=\SI{550}{nm}$) decreases down to $2\%$ of its original value~\cite{ITU2}. The formula we adopted is the empirical one typically applied in case of fog~\cite{book:kaushal2017,ITU1,ITU2}:
\begin{eqnarray}\label{Mie_scattering}
& \beta_{\mathrm{a}}(\lambda) = 10 \log\left(\mathrm{e}\right)  \frac{3.91}{V} \left(\frac{\lambda}{\lambda_0}\right)^{-p}
\\  \mathrm{with } \ \ \ \ \ & p = \begin{cases}
1.6                     &  V > \SI{50}{km}\\
1.3                     &  \SI{6}{km}<V<\SI{50}{km} \\
0.585 V^{\frac{1}{3}}   &  V<\SI{6}{km}
\end{cases}
\end{eqnarray}
where $\lambda$ is expressed in~nm, and $p$ is a coefficient related to the size distribution of the scattering particles, according to the Kruse model~\cite{book:kaushal2017}. Starting from this empirical formula, it is possible to evaluate the attenuation due to weather conditions in several cases such as heavy fog, light haze/drizzle and clear sky~\cite{book:kaushal2017}. In case of intense rain or snow, which are outside the purpose of this work, other formulas must be considered~\cite{ITU2,ITU1}. 

The effect of turbulence, $ A_{\textrm{sci}} (dB)$ can be taken into account by using the formula~\cite{MALIK:2020,HANDURA2016}:
\begin{equation}
    A_{\textrm{sci}} \mathrm{(dB)} = 2 \cdot \sqrt{ 23.17 \cdot k^{7/6} \cdot C_{\textrm{n}}^2 \cdot L^{11/6}}
\end{equation}
that describes the losses due to scintillation, where $k=2\pi / \lambda$ is the wavenumber, $L$ is the link range in meter and $C_{\textrm{n}}$ is the refractive index structure parameter in m$^{2/3}$ calculated via the Hufnagel-Valley model \cite{valley1980isoplanatic}. A wind speed of \SI{30}{km/h} and a quote of \SI{50}{m} over the sea level are considered to retrieve the $C^2_{\mathrm{n}}$ factor in the typical case of moderate turbulence condition~\cite{MALIK:2020}.
Our simulation also accounts for the geometrical attenuation factor due to the Gaussian propagation of the laser beam, leading to a divergence in the far-field region.
The geometrical attenuation results in a $1/d^2$ scaling of the far-field intensity impinging on the detector, where $d$ represents the TX-RX distance. Furthermore, it depends on the laser wavelength and on the optical aperture of the light-collecting system at the receiver side~\cite{ITU1,ITU2}. In this work, we simulate a system where the radius of both the transmitter and the receiver aperture is \SI{10}{cm}. We estimate the geometrical attenuation coefficient $A_{\mathrm{geo}}$ via the following equation~\cite{ITU1,ITU2}:
\begin{equation}
A_{\mathrm{geo}}(dB)= \begin{cases} 10 \log \left(\dfrac{S_{\mathrm{d}}}{S_{\mathrm{capture}}}\right) & \text{$S_{\textrm{d}}>S_{\mathrm{capture}}$} , \\
0 & \text{otherwise} ,
\end{cases}
\end{equation}
where $S_{\mathrm{d}}$ is the wavefront area of the transmitted beam at the receiver at a distance $d$, and $S_{\mathrm{capture}}$ is the receiver capture surface.
Within short distances it is possible that $S_{\mathrm{capture}}$ is larger than the beam area. In this case, all the light is collected and $A_{\mathrm{geo}}(dB)$ is equal to zero~\cite{ITU1,ITU2}. For sake of simplicity, in our model we assume this to happen for distances lower than twice the Rayleigh length, where we assume no geometrical losses. For longer distances we evaluate the losses considering a receiver aperture smaller than the diameter of the diverging beam.
In Fig.~\ref{fig:fig3}, we show the combined atmospheric and the geometrical attenuation coefficients for two different values of visibility $V$, making a comparison between the wavelength used in this experiment ($\lambda=\SI{4720.0}{nm}$), with the optimal Mid-IR wavelength for air transmission ($\lambda=\SI{3998.6}{nm}$) and the optimal telecom one ($\lambda=\SI{1557.7}{nm}$). 
The total attenuation $A_\mathrm{all} (dB)$ reported in Figure is calculated as $A_\mathrm{all} (dB) = A_{\mathrm{geo}}(dB) + A_{\textrm{sci}} (dB) + \gamma(dB)
$, with $\gamma (dB)$ is $\gamma(\lambda)$ expressed in dB.
In particular, the optimal Mid-IR wavelength around \SI{4}{\micro \meter} features the lowest absorption as a result of a thorough high-resolution analysis of the atmospheric absorption spectrum provided by the HITRAN database~\cite{hitrandat}. Assuming the same setup geometry, the impact of geometrical attenuation is in general greater in the Mid-IR than in the NIR due to the larger wavelength. On the other hand, scintillation effects impact more on the telecom wavelengths. In the case of very clear air condition, corresponding to $V=\SI{23}{km}$~\cite{ITU1,book:kaushal2017}, the top plot in Fig.~\ref{fig:fig3} shows that the optimal Mid-IR wavelength (orange dashed curve) is less attenuated than the other wavelengths in all the distance range took into account. Over short distances, below 10 km, the used wavelength (green curve) is still convenient over the NIR one (blue curve), while for longer distances the two wavelengths perform similarly. On the other hand, the lower graph shows the expected optical channel attenuation as a function of distance in case of low visibility, $V=\SI{1}{km}$, corresponding to heavy fog and cloud~\cite{book:kaushal2017}, which is dominated by scattering. Remarkably, in this case of low visibility, the Mid-IR wavelengths are in general much less affected by the losses than NIR ones, and the optimal system at \SI{4.0}{\micro m} \cite{sauvage_outdoor_MIR_link_noise} shows an advantage of 4--5~dB over the whole explored distance range of \SI{20}{\kilo \meter} for low visibility. Interestingly, however, Fig. \ref{fig:fig3} also highlights that in the atmospheric conditions set for the simulation (moderate turbulence), the optimal Mid-IR systems outperforms the standard telecom one also in the large visibility condition, due to an optimal combination of geometrical propagation and reduced scattering properties.

\section{Experimental Results and discussion}
\label{sec:result&di}
As anticipated in Sec.~\ref{sec:theory}, in the following Section we characterize the system performance in terms of PER and SNR recorded in  both HAR and LAR configurations. 

Our measurements will then be combined with the predictions of the channel model discussed in Sec. \ref{sec:theory} to give an estimation of the maximum Mid-IR link length attainable with our system in various realistic visibility conditions. 
\subsection{High Attenuation Regime (HAR)}
\label{HAR}
\begin{figure}[ht!]
\centering
\includegraphics[scale=0.5]{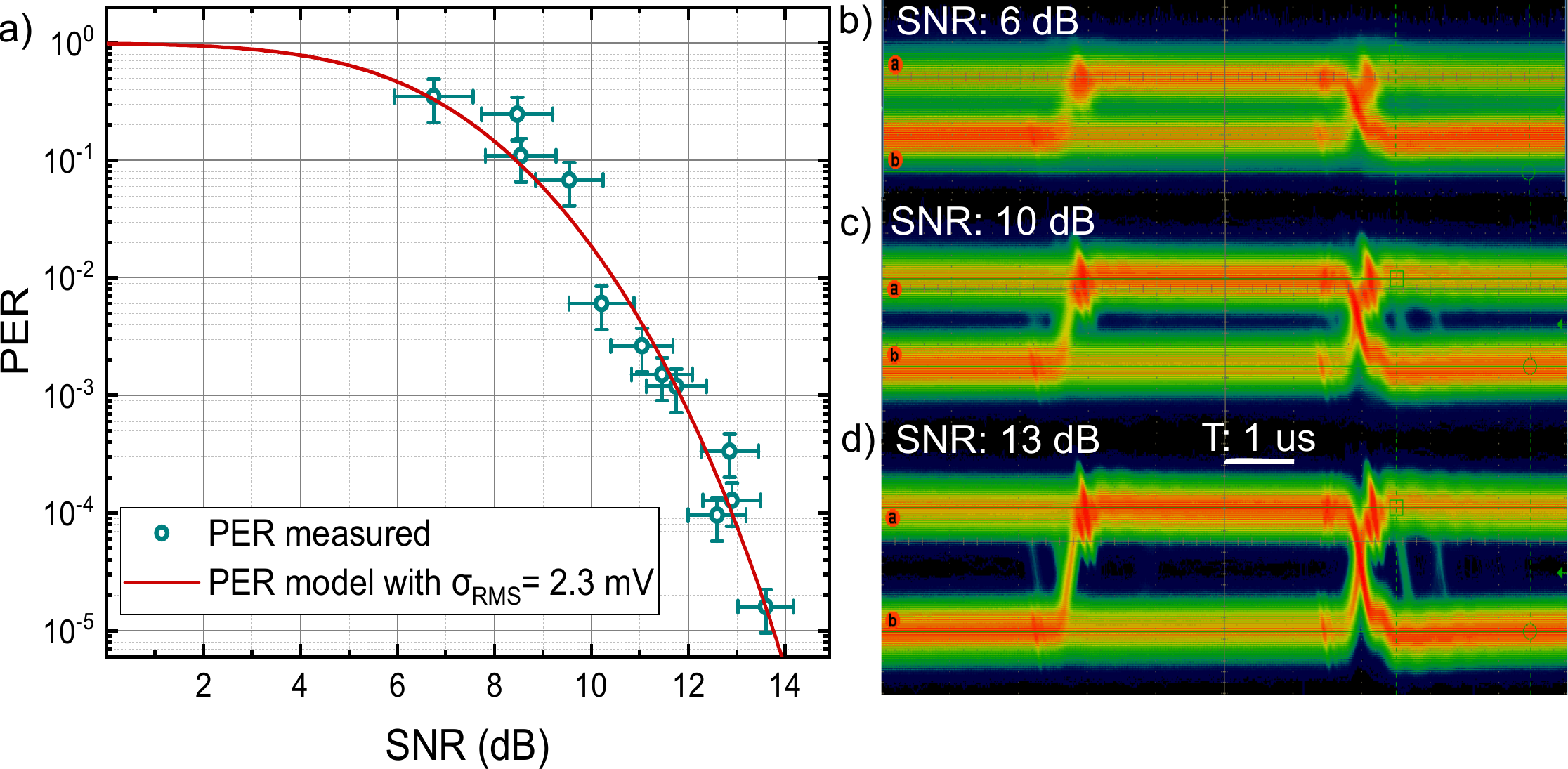}
\caption{a) PER as a function of the SNR in the HAR. Measured data points (green circles) feature a very nice agreement with predictions of the PER model for a $\sigma_{\mathrm{RMS}}$ = \SI{2.3}{mV}, which corresponds to the measured RMS. The error-free communication threshold of $1.6 \times 10^{-5}$ is achieved for SNR > \SI{14}{dB}. The vertical error bars orrespond to the standard deviation after repeated measurements, while horizontal error bars are obtained after error propagation from measurements of $\sigma_{\mathrm{RMS}}$ and S$_{RX}$ values.
b),c) and d) Eye patterns for three different values of SNR (\SI{6}, \SI{10}, \SI{13}{dB} respectively) corresponding to three different PER regimes. The horizontal scale is \SI{1}{\micro \second/ div}, while the vertical scale is \SI{10}{mV/ div}. The acquisition is self triggered on the received signal. \label{fig:fig4}}
\end{figure}
\begin{figure}[ht!]
\centering
\includegraphics[width=0.6\linewidth]{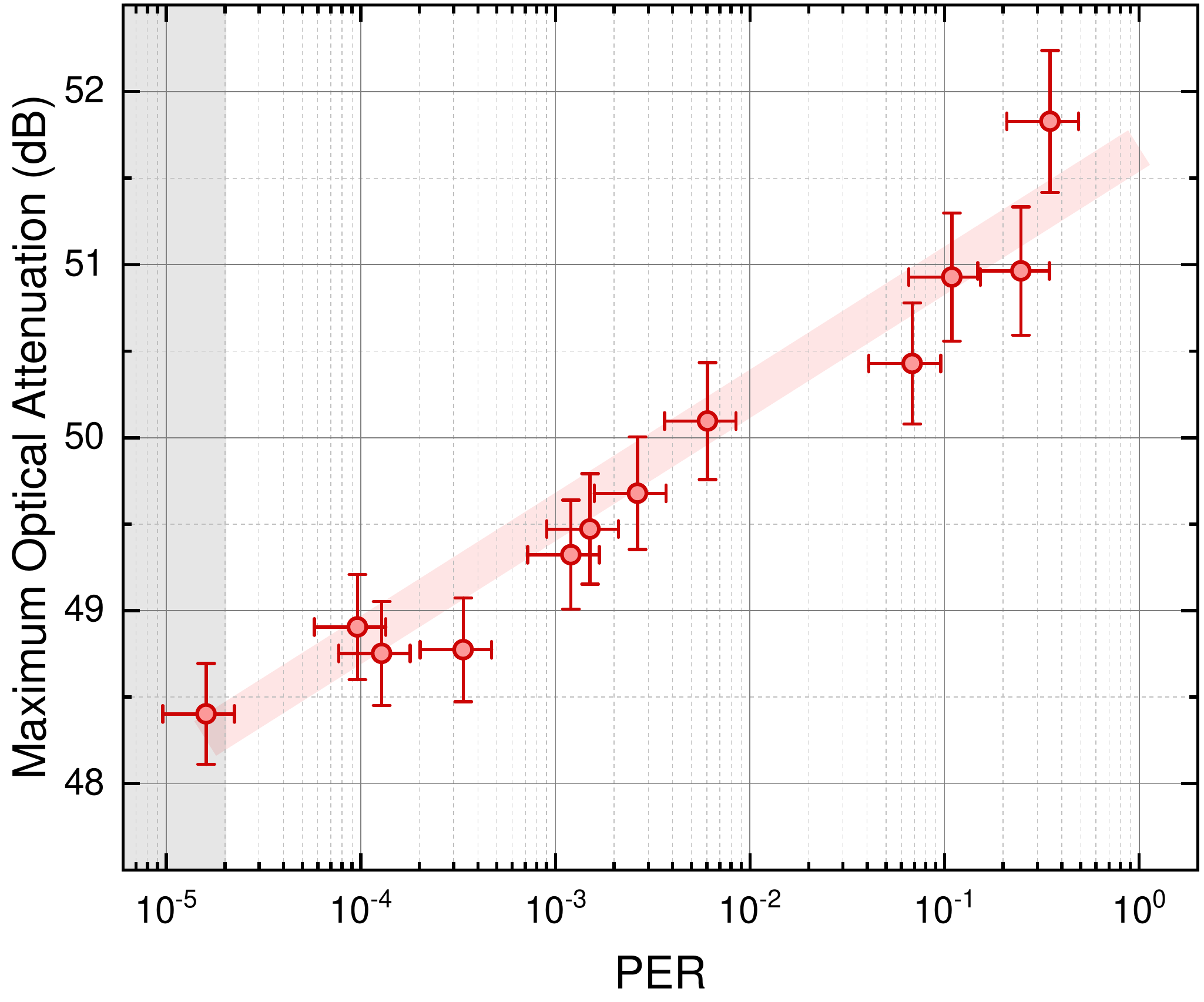}
\caption{MOA as a function of the PER for MD=100\% in the HAR. The gray shaded area highlights the threshold for an error-free communication, it is achieved for MOA lower than \SI{48}{dB}. The oblique shaded stripe is a guide to the eye, to highlight the trend of MOA. MOA increases by \SI{2}{dB} every 3 decades of PER. 
}\label{fig:fig5}
\end{figure}
In the HAR we aim at determining the system response as a function of channel attenuation, and to give an estimation of the MOA tolerable by our QCL-based Mid-IR communication system for granting reliable optical links given a requested PER value.
In Fig.~\ref{fig:fig4}a), we first show the dependence of the measured PER on the SNR, considering  the recorded value of $\sigma_{\mathrm{RMS}} = \SI{2.3}{mV}$ given by the detector background noise.
The error bars on PER are obtained as the standard deviation on repeated measurements, while the SNR error bars are obtained after error propagation from measurements of $\sigma_{\mathrm{RMS}}$ and S$_{Rx}$ values.
The red curve represents the PER model as a function of SNR (see Sec. \ref{sec:theory}), fixing  $\sigma_{\mathrm{RMS}}$ at \SI{2.3}{mV}. It is in good agreement with data.
The error-free communication (PER < 1.6 $\times$ 10$^{-5}$) is achieved for a SNR larger than \SI{14}{dB}. 
Fig.~\ref{fig:fig4} also shows the recorded eye patterns in  low-signal (SNR = \SI{6}{dB}, PER $\sim$ 0.5, panel b)), medium-signal (SNR = \SI{10}{dB}, PER $\sim$ 0.02, panel c)), and high-signal (SNR = \SI{13.5}{dB}, error-free, panel d)) configurations. The traces report the self-triggered signal after the amplified photodetector RX stage. 
The jitter observed on the transition edges of the eye pattern depends on the signal quality and its minimum value is due to the time resolution of the Arduino DUE platform.
Fig.~\ref{fig:fig4} reports the observed MOA as a function of PER, for MD=100\%. An error-free communication is achieved for MOA lower than \SI{48}{dB}. Assuming an internet connection reference value of PER = $10^{-3}$, the relative observed MOA is slightly higher (49.5 dB). The error bars on MOA are calculated  with the  propagation of the statistical errors obtained during signal and noise acquisitions. The shaded stripe in Fig.~\ref{fig:fig4} is a  guide to the eyes and suggests that the MOA increases by \SI{2}{dB} every 3 decades of PER (for PER$<1$). This behaviour agrees with the expected theoretical trend described in Sec.~\ref{sec:theory}. For example, let's consider two different values of MOA$=48.5$ and MOA$=50.5$ for MD = 100\%. The relative values of SNR(dB) obtained from~\eqref{eq:combined} are  SNR = \SI{13}{dB} (MOA$=48.5$) and SNR = \SI{9}{dB} (MOA=$50.5$). 
By converting these values in linear scale, through the relations reported in Sec. \ref{sec:theory}) 
we can obtain the relative PERs values. In agreement with our observations, these differ by 3 decades for the two selected values of MOA, approximately.
\subsection{Low attenuation regime (LAR)}
\begin{figure}[t!]
\centering
\includegraphics[scale=0.5]{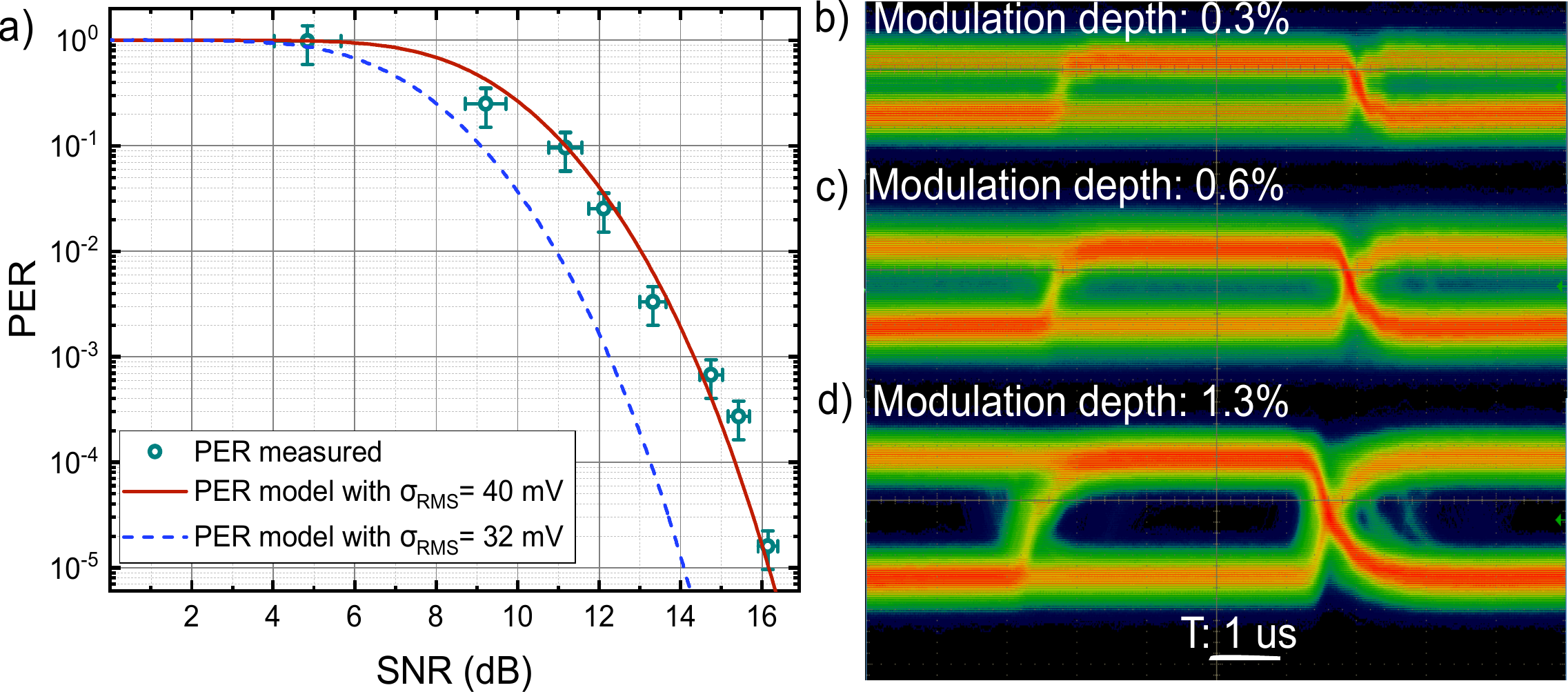}
\caption{a) Communication performance as a function of the SNR in the LAR. The green circles represent the measured PER, while the dashed blue curve and the red solid curve correspond to the simulated PER with $\sigma_{\mathrm{RMS}} $ = \SI{32}{mV} (experimental value) and $\sigma_{\mathrm{RMS}} $ = \SI{40}{mV} (best agreement value), respectively. b)-d) Eye patterns for three different MDs: 0.3\% (corresponding to PER $\sim$ 0.2); 0.6\% (corresponding to PER $\sim$ 0.003); 1.3\% (error-free communication).
}\label{LAR_SNR}
\end{figure}
\begin{figure}[t!]
\centering
\includegraphics[width=0.6\columnwidth]{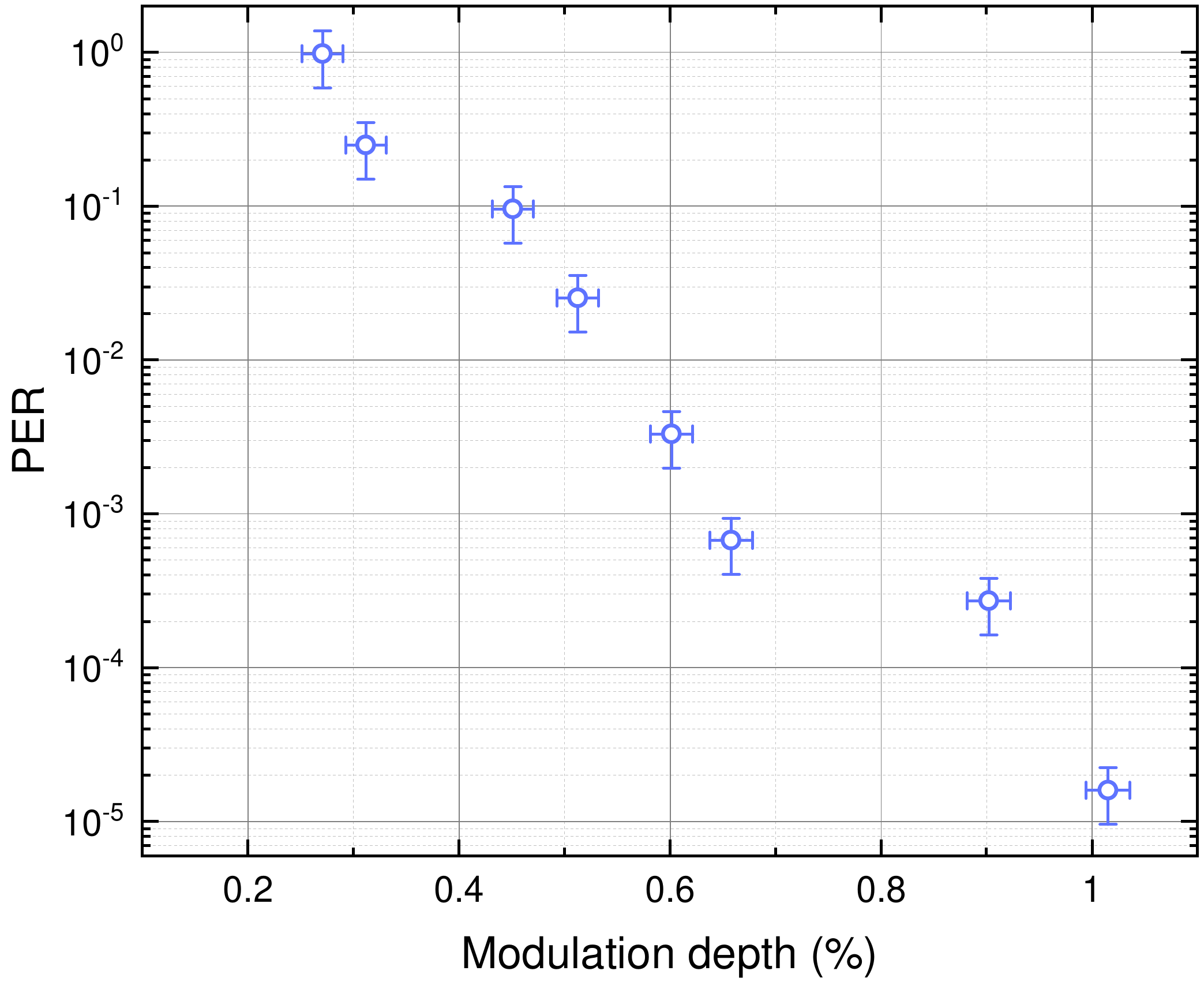}
\caption{PER as a function of the MD in the LAR, for a fixed OA of 13 dB. The error-free communication is established for MDs greater than 1\%.}\label{LAR depth}
\end{figure}
In the LAR, both noise and signal amplitude are expected to grow linearly with $P_{out}$. In contrast, however, SNR is still expected to grow with MD. In particular, for MDs much lower than 100\%, we expect SNR $\propto$ MD. For this reason, we investigate the minimum MD (in the low-modulation regime) needed for an error-free communication for given values of OA in the optical link. By identifying the minimum required modulation, we can define the best working condition for the setup in terms of laser stability and spectral quality. This is a relevant issue when, e.g., more sophisticated modulation schemes (such as OFDM) aimed at larger bit-rates are involved, where both amplitude and frequency stability of the baseband signal are critical factors. In such cases, a strict single-mode operation of the laser source is essential.
As in the case of HAR, we first evaluate the communication performance against the  recorded SNR (Fig.~\ref{LAR_SNR}). In this case, the PER model curve which most accurately describes the measured values (green circles) is obtained by choosing a $\sigma_{\mathrm{RMS}} $ = \SI{40}{mV} (red curve). This $\sigma_{\mathrm{RMS}} $ value slightly differs from the measured RMS noise value $\sigma_{\mathrm{RMS}} $ = \SI{32}{mV} (blue dashed line), which we measure when no transmission occurs. This slight underestimation of noise value could be addressed to the presence of fast transients and glitches that can occur during the transmission, as a consequence of irradiated EM noise due to the large and steep variations of current levels involved in the modulation process.
We note that in the HAR this effect is negligible, as the predominant noise  contribution is related to the background noise of the detector.
Fig.~\ref{LAR_SNR} also shows the eye patterns for three different MDs (0.3 \% panel b), 0.6 \% panel c), and 1.3 \% panel d)) corresponding to low, medium and high SNR (SNR=$\SI{9}{dB}$ with a PER $\sim 0.2$, SNR=$\SI{13}{dB}$ with PER $\sim 0.003$, and SNR$>\SI{16}{dB}$ with PER$\lesssim 10^{-5}$, respectively).
As introduced in Sec.~\ref{sec:theory}, in case of LAR we experimentally evaluate the communication performance of our FSOCS for different MDs, to find the minimum MD required to perform an error-free communication for a given value of P$_{\mathrm{out}}$. We remark that in this regime, where the detection noise is dominated by the intrinsic intensity noise of the source, we do not expect significant improvement in the communication quality by increasing the optical power emitted by the QCL source. The SNR(dB) linearly improves with the MD as described by Equation \eqref{eq:combined} and, with reference to Fig.~\ref{LAR depth}, an error-free communication is obtained for MDs greater than 1\%. The error bars are calculated as in the HAR case. 
\begin{figure}[ht]
\centering
\includegraphics[width=0.7\linewidth]{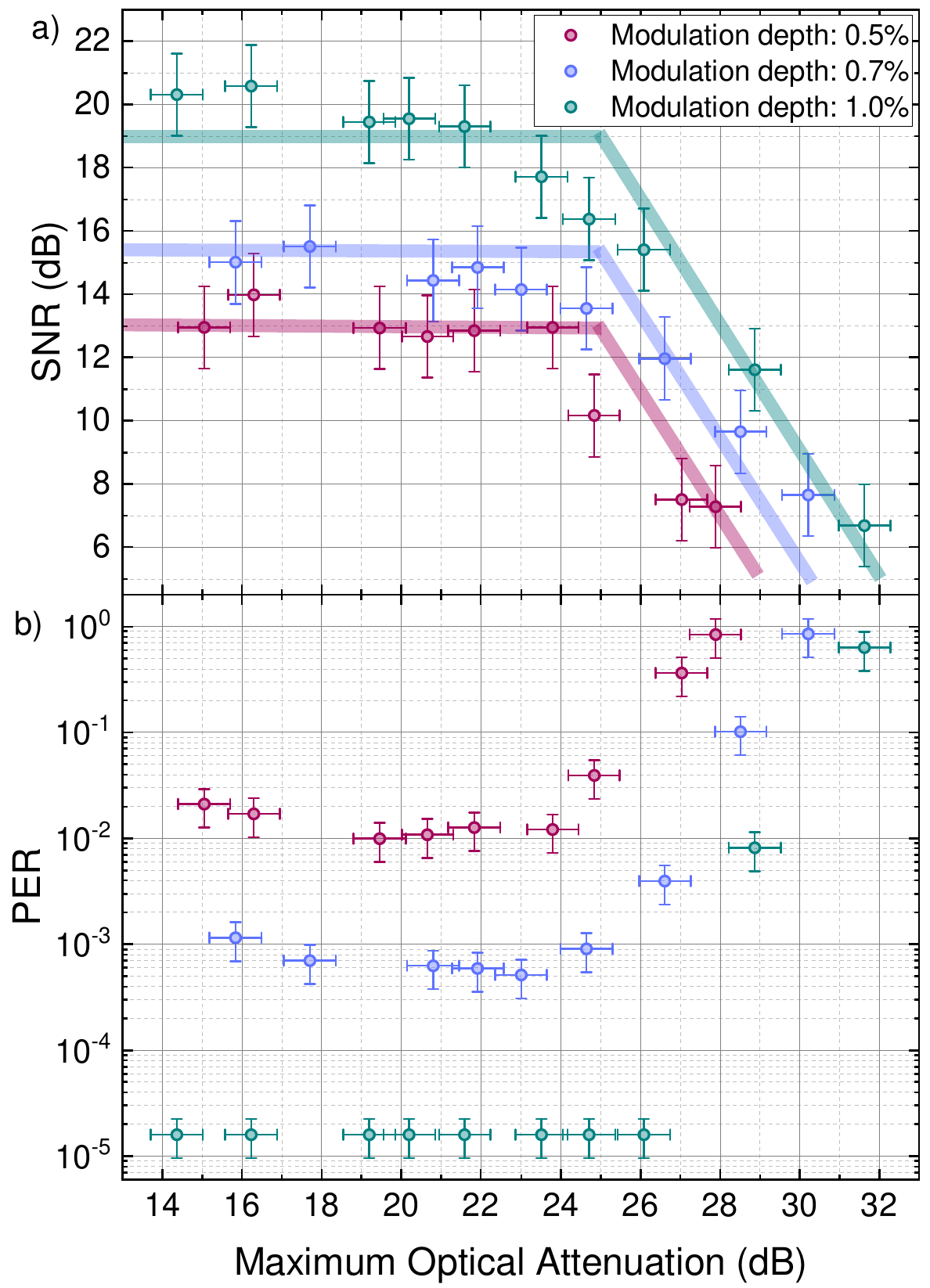}
\caption{a) SNR as a function of the MOA for three different MDs, 0.5\% (purple dots), 0.7\% (light blue dots) and 1\% (green dots). For MOA up to $\sim$ \SI{25}{dB}, the system is limited by the QCL noise, which increases as the signal, therefore the SNR remains constant (shaded areas, which are only a guide to the eye). For greater attenuation, the system is limited by the detector noise floor and the SNR approximately decreases by \SI{2}{dB/dB} (slanting lines). b) PER as a function of MOA across the HAR-LAR transition region. The PER shows a plateau for low MOA values, and increases for larger MOAs, as suggested by Equation \eqref{eq:combined}.} 
\label{HAR-LAR pics}
\end{figure}
\subsection{HAR-LAR transition}
The versatility of our FSOCS, along with the testing facility, allowed us to investigate the behaviour of the system in the transition between HAR and LAR, i.e. between the noise regimes described in Sec.\ref{sec:noise}. In particular, we vary the attenuation of the optical link from \SI{15}{dB} to \SI{32}{dB} for three different MDs (0.5\%, 0.7\% and 1\%), as shown in Fig.~\ref{HAR-LAR pics}.
Even in this case the laser operates in single-mode regime well above threshold ($I=\SI{663}{mA}$,  $P_{\mathrm{out}}=\SI{13.6}{mW}$).

Fig.~\ref{HAR-LAR pics}a) reports the observed SNR as a function of the MOA. Our observations clearly confirm the existence of two distinct noise regimes depending on the global attenuation affecting the optical channel. For low attenuations (LAR), where the detection noise is dominated by the QCL intensity noise, SNR does not significantly depend on the optical signal impinging on the RX stage. In contrast, for MOA larger than a critical value (\SI{25}{dB} in our case) the SNR decreases as the attenuation grows (HAR). The shaded areas, intended as a guide to the eye, highlight a SNR vs MOA decrease ratio of $\sim$ \SI{2}{dB/dB}, which is compatible with predictions of Equation \eqref{eq:combined} in the HAR regime. In the LAR, the constant value of the SNR (given by k$_i$ in Equation \eqref{eq:combined}), depends on the MD, as this influences the effective amplitude of AC signal recorded by the RX stage. The reduction in the (constant) SNR values recorded in LAR for different MDs are in good agreement with the expectations yielded by Equation \eqref{eq:combined}. For example, with reference to Fig.~\ref{HAR-LAR pics}a), a reduction of 50\% in the MD results in an effective reduction of \SI{-6}{dB} in the observed SNR, as a comparison between green and red data points confirms.
The observed behaviour is reflected into the communication performance, where a steep increase in the recorded PER is observed above the \SI{25}{dB} transition point (HAR). Conversely, in agreement with what has been observed for the SNR values, in the LAR the system is featuring stable communication performance independently on the channel attenuation, as the main noise source in the detection stage is represented by the intrinsic intensity noise of the QCL source.
%
\subsection{Estimation of the outdoor performance of the FSO links}
We now relate our experimental findings on our Mid-IR FSOCS to possible realistic scenarios. In particular, we are interested in estimating the maximum effective length of an FSO link employing our communication system under adverse atmospheric conditions, where Mid-IR links are expected to outperform NIR ones. We remark that whilst such estimation does not exactly correspond to real FSO link lengths as scintillation is the only turbulence phenomenon accounted for by our model, such estimation still provides very realistic insights on the potential cast of our Mid-IR FSOCS as compared to other systems in real conditions. To obtain the effective link lengths, we consider the case of large attenuation (HAR) and maximal MD (100\%). We can combine our experimental results on MOA in HAR (Fig.~\ref{fig:fig5}) with the expected channel attenuation given by the model predictions (Fig.~\ref{fig:fig3}). According to Fig.~\ref{fig:fig5}, an error-free communication (PER$<1.6\times10^{-5}$) requires MOA < \SI{48}{dB}. A comparison with Fig.~\ref{fig:fig3}, allows us to obtain the error-free communication distance in both high- and low- visibility cases ($V = \SI{1}{km}$ and $V = \SI{23}{km}$), and for the three values of wavelength discussed in Sec. \ref{sec:theory} and Sec.\ref{sec:noise} (i.e., a telecom NIR source ($\lambda\sim$  \SI{1.56}{\micro m}), our Mid-IR source ($\lambda\sim$  \SI{4.72}{\micro m}), and the optimal Mid-IR wavelength ($\lambda \sim$  \SI{4.0}{\micro m}) which would minimize the effects of atmospheric absorption and scattering). 
\begin{table}[htbp!]
\begin{center}
{\sffamily  \begin{tabular}{ | m{1.5cm} | m{1.8cm}| m{1.8cm} | m{1.8cm} |}
\hline
&  \bf{NIR source} \newline ($\lambda$ = \SI{1.56}{\micro m}) & \bf{Mid-IR source} \newline ($\lambda$ = \SI{4.72}{\micro m})  & \bf{Mid-IR source}  \newline ($\lambda$ = \SI{4.0}{\micro m})  \\ \hline
\bf{V = \SI{1}{km}} & < \SI{4.5}{km} & < \SI{8.0}{km} & < \SI{7.8}{km}\\ \hline
\end{tabular}}
\end{center}
\caption{Estimated effective link lengths for an error-free communication in low-visibility conditions (V = \SI{1}{km}) 
for a conventional telecom source ($\lambda$ = \SI{1.56}{\micro m}), and for two different Mid-IR sources (the one used in this experiment at \SI{4.72}{\micro m} and one with the optimal transmission wavelength around \SI{4.0}{\micro m}). We consider a MD=100\% and HAR configuration.}\label{table:PER distances}
\end{table}

Table ~\ref{table:PER distances} reports the expected maximum link lengths estimated through such analysis.
In the low-visibility case, our Mid-IR prototype at \SI{4.72}{\micro m} is expected to grant error-free communication for link lengths up to \SI{8.0}{km}, larger than the attainable distance of \SI{4.5}{km} estimated for the standard NIR telecom source at \SI{1.56}{\micro m}. The performance at \SI{4.0}{\micro m} wavelength is comparable to the one at \SI{4.72}{\micro m}. This effect is mostly due to the reduced scattering effects, which are larger for short wavelengths (Eq.~\ref{Mie_scattering}), confirming the extreme relevance of Mid-IR FSO links as valid alternative to NIR telecom systems in case of adverse atmospheric conditions. 
We remark that the \SI{4.0}{\micro m} range is at full reach of actual Mid-IR QCL chips, making QCLs one of the most versatile platform for optimal FSOCS to be employed in realistic applications.

\section{Conclusions} 
\label{sec:conclusions}
In this work, we have presented for the first time an extensive characterization of noise regimes that could occur in QCL-based FSOCS, based on a versatile Mid-IR QCL system at \SI{4.72}{\micro m}. We carried out a detailed study of communication performances in two different noise regimes (HAR and LAR), finding a very different response of the system in terms of transmission quality as a function of the optical channel attenuation. In the HAR, where the predominant noise contribution is given by the detector noise, the system communication is tested against the maximal optical attenuation tolerable in order to achieve error-free communication (PER$<1.6\times 10^{-5}$), finding MOA values as high as 48 dB for 100\% modulation depths and a baudrate of 115 kbaud. In contrast, in the LAR regime, which is more typical of short-range links, we observed an almost constant PER as a function of the optical attenuation, as a consequence of a SNR value which is independent of the received signal amplitude $S_{RX}$. 
The versatility of our setup also allowed us to characterize the transition region between the HAR and LAR regimes, in terms of both SNR and tolerable MOA, finding a clear crossing point between the two regimes. 
We also estimate the performance of our Mid-IR FSOCS under realistic operational conditions by combining our findings with the predictions of a simplified propagation model taking into account both geometrical and atmospheric (absorption, scattering and scintillation) effects for moderate turbulence, comparing them with the performance expected for different NIR and Mid-IR wavelengths. The estimated error-free link length for the presented Mid-IR FSOCS is \SI{8.0}{\kilo \meter} in low visibility conditions ($V=1$~km). Noticeably enough, in such a low visibility condition, this overwhelms the expected performance for a standard source in the telecom range which, in contrast, is favored by a lower divergence in the far field due to the shorter wavelength. However, our analysis shows that a FSOCS based on Mid-IR QCLs at an optimal wavelength of \SI{4.0}{\micro m} \cite{sauvage_outdoor_MIR_link_noise} could also overperform NIR systems in good visibility, simultaneously featuring excellent resilience to scattering and good propagation properties.  

The results presented in this work have a general breadth, and could have a deep impact for future QCL-based FSOCs, also working in different wavelength regions and featuring Gbps-class bitrates. 

%

\section*{Funding}

The Authors acknowledge financial support from the European Union’s Horizon 2020 Research and Innovation Programme (Qombs Project, FET Flagship on Quantum Technologies grant n. 820419; QuaLIDAD Project, FET Flagship on Quantum Technologies - FET Innovation Launchpad grant n. 101034794; Laserlab-Europe Project grant n. 871124), from the Italian ESFRI Roadmap (Extreme Light Infrastructure - ELI Project), from the projects MIUR PON 2017 ARS01\_00917 ”OK-INSAID”, and MIUR FOE Progetto Premiale 2015 ”OpenLab 2”. 

\section*{Acknowledgments}
The Authors gratefully thank the collaborators within the consortium of the Qombs Project: Prof. Dr. Jérome Faist (ETH Zurich) for having provided the quantum cascade laser and the company ppqSense for having provided the ultra-low-noise current driver (QubeCL). In addition, the Authors gratefully thank all the members of the VisiCore joint laboratory for Research on Visible Light Communications. 

\section*{Disclosures} The authors declare no conflicts of interest.


\bibliography{refs}

\begin{thebibliography}{10}
\newcommand{\enquote}[1]{``#1''}

\bibitem{Willebrand:2001}
H.~Willebrand and B.~Ghuman, \enquote{Fiber optics without fiber,}
  {\protect\JournalTitle{IEEE Spectrum}} \textbf{38}, 40--45 (2001).

\bibitem{esmail:2019}
M.~A. Esmail, A.~M. Ragheb, H.~A. Fathallah, M.~Altamimi, and S.~A. Alshebeili,
  \enquote{5g-28 ghz signal transmission over hybrid all-optical fso/rf link in
  dusty weather conditions,} {\protect\JournalTitle{IEEE Access}} \textbf{7},
  24404--24410 (2019).

\bibitem{Mirabissi}
D.~Marabissi, L.~Mucchi, S.~Caputo, F.~Nizzi, T.~Pecorella, R.~Fantacci,
  T.~Nawaz, M.~Seminara, and J.~Catani, \enquote{Experimental measurements of a
  joint 5g-vlc communication for future vehicular networks,}
  {\protect\JournalTitle{Journal of Sensor and Actuator Networks}} \textbf{9}
  (2020).

\bibitem{Klaus:2018}
K.~David and H.~Berndt, \enquote{{6G} vision and requirements: Is there any
  need for beyond {5G}?} {\protect\JournalTitle{IEEE Vehicular Technology
  Magazine}} \textbf{13}, 72--80 (2018).

\bibitem{Dang:2020}
S.~Dang, O.~Amin, B.~Shihada, and M.-S. Alouini, \enquote{What should 6g be?}
  {\protect\JournalTitle{Nature Electronics}} \textbf{3}, 20--29 (2020).

\bibitem{Dehghani:2021}
M.~Dehghani~Soltani, E.~Sarbazi, N.~Bamiedakis, P.~de~Souza, H.~Kazemi, R.~V.
  Penty, H.~Haas, and M.~Safari, \enquote{Safety analysis for laser-based
  optical wireless communications: A tutorial,} {\protect\JournalTitle{arXiv
  e-prints}} pp. arXiv--2102 (2021).

\bibitem{Khan:2017}
M.~T.~A. Khan, M.~A. Shemis, E.~Alkhazraji, A.~M. Ragheb, M.~A. Esmail,
  H.~Fathallah, S.~Alshebeili, and M.~Z.~M. Khan, \enquote{Optical wireless
  communication at 100 {Gb/s} using l-band quantum-dash laser,} in \emph{2017
  Conference on Lasers and Electro-Optics Pacific Rim (CLEO-PR),}  (2017), pp.
  1--3.

\bibitem{Bloom:2003}
S.~Bloom, E.~Korevaar, J.~Schuster, and H.~Willebrand, \enquote{Understanding
  the performance of free-space optics $invited$,} {\protect\JournalTitle{J.
  Opt. Netw.}} \textbf{2}, 178--200 (2003).

\bibitem{faist:1994}
J.~Faist, F.~Capasso, D.~L. Sivco, C.~Sirtori, A.~L. Hutchinson, and A.~Y. Cho,
  \enquote{Quantum cascade laser,} {\protect\JournalTitle{Science}}
  \textbf{264}, 553--556 (1994).

\bibitem{Su:18}
Y.~Su, W.~Wang, X.~Hu, H.~Hu, X.~Huang, Y.~Wang, J.~Si, X.~Xie, B.~Han,
  H.~Feng, Q.~Hao, G.~Zhu, T.~Duan, and W.~Zhao, \enquote{10 {Gbps} {DPSK}
  transmission over free-space link in the mid-infrared,}
  {\protect\JournalTitle{Opt. Express}} \textbf{26}, 34515--34528 (2018).

\bibitem{Corrigan:2009}
P.~Corrigan, R.~Martini, E.~A. Whittaker, and C.~Bethea, \enquote{Quantum
  cascade lasers and the kruse model in free space optical communication,}
  {\protect\JournalTitle{Opt. Express}} \textbf{17}, 4355--4359 (2009).

\bibitem{Flannigan_2022}
L.~Flannigan, L.~Yoell, and C.~qing Xu, \enquote{Mid-wave and long-wave
  infrared transmitters and detectors for optical satellite
  communications{\textemdash}a review,} {\protect\JournalTitle{Journal of
  Optics}} \textbf{24}, 043002 (2022).

\bibitem{Tombez:2013a}
L.~Tombez, F.~Cappelli, S.~Schilt, G.~Di~Domenico, S.~Bartalini, and
  D.~Hofstetter, \enquote{Wavelength tuning and thermal dynamics of
  continuous-wave mid-infrared distributed feedback quantum cascade lasers,}
  {\protect\JournalTitle{Appl. Phys. Lett.}} \textbf{103}, 031111 (2013).

\bibitem{faist:2013}
J.~Faist, \emph{Quantum cascade lasers} (OUP Oxford, 2013).

\bibitem{Riedi:2015}
S.~Riedi, F.~Cappelli, S.~Blaser, P.~Baroni, A.~M\"{u}ller, and J.~Faist,
  \enquote{Broadband superluminescence, $5.9~\mu$m to $7.2~\mu$m, of a quantum
  cascade gain device,} {\protect\JournalTitle{Opt. Express}} \textbf{23},
  7184--7189 (2015).

\bibitem{cathabard2010quantum}
O.~Cathabard, R.~Teissier, J.~Devenson, J.~Moreno, and A.~Baranov,
  \enquote{Quantum cascade lasers emitting near 2.6 $\mu$m,}
  {\protect\JournalTitle{Applied Physics Letters}} \textbf{96}, 141110 (2010).

\bibitem{paiella:2001}
R.~Paiella, R.~Martini, F.~Capasso, C.~Gmachl, H.~Y. Hwang, D.~L. Sivco, J.~N.
  Baillargeon, A.~Y. Cho, E.~A. Whittaker, and H.~Liu, \enquote{High-frequency
  modulation without the relaxation oscillation resonance in quantum cascade
  lasers,} {\protect\JournalTitle{Applied Physics Letters}} \textbf{79},
  2526--2528 (2001).

\bibitem{hinkov:2016}
B.~Hinkov, A.~Hugi, M.~Beck, and J.~Faist, \enquote{Rf-modulation of
  mid-infrared distributed feedback quantum cascade lasers,}
  {\protect\JournalTitle{Optics express}} \textbf{24}, 3294--3312 (2016).

\bibitem{Consolino:2018a}
L.~Consolino, F.~Cappelli, M.~Siciliani~de Cumis, and P.~De~Natale,
  \enquote{{QCL}-based frequency metrology from the mid-infrared to the {THz}
  range: a review,} {\protect\JournalTitle{Nanophotonics}} \textbf{8}, 181--204
  (2018).

\bibitem{borri:2019}
S.~Borri, G.~Insero, G.~Santambrogio, D.~Mazzotti, F.~Cappelli, I.~Galli,
  G.~Galzerano, M.~Marangoni, P.~Laporta, V.~Di~Sarno \emph{et~al.},
  \enquote{High-precision molecular spectroscopy in the mid-infrared using
  quantum cascade lasers,} {\protect\JournalTitle{Applied Physics B}}
  \textbf{125}, 18 (2019).

\bibitem{gutowska:2011}
M.~Gutowska, D.~Pier{\'s}ci{\'n}ska, M.~Nowakowski, K.~Pierci{\'n}ski,
  D.~Szabra, J.~Miko{\l}ajczyk, J.~Wojtas, and Z.~Bielecki,
  \enquote{Transmitter with quantum cascade laser for free space optics
  communication system,} {\protect\JournalTitle{Bulletin of the Polish Academy
  of Sciences. Technical Sciences}} \textbf{59}, 419--423 (2011).

\bibitem{mikolajczyk:2014}
J.~Miko${\l}$ajczyk, \enquote{An overview of free space optics with quantum
  cascade lasers,} {\protect\JournalTitle{International Journal of Electronics
  and Telecommunications}} \textbf{60}, 259--264 (2014).

\bibitem{Liu:2015}
C.~Liu, S.~Zhai, J.~Zhang, Y.~Zhou, Z.~Jia, F.~Liu, and Z.~Wang,
  \enquote{Free-space communication based on quantum cascade laser,}
  {\protect\JournalTitle{Journal of Semiconductors}} \textbf{36}, 094009
  (2015).

\bibitem{Corrias:2022}
N.~Corrias, T.~Gabbrielli, P.~De~Natale, L.~Consolino, and F.~Cappelli,
  \enquote{Analog {FM} free-space optical communication based on a mid-infrared
  quantum cascade laser frequency comb,} {\protect\JournalTitle{Opt. Express}}
  \textbf{30}, 10217--10228 (2022).

\bibitem{spitz2022}
O.~Spitz, P.~Didier, L.~Durupt, D.~A. Díaz-Thomas, A.~N. Baranov, L.~Cerutti,
  and F.~Grillot, \enquote{Free-space communication with directly modulated
  mid-infrared quantum cascade devices,} {\protect\JournalTitle{IEEE Journal of
  Selected Topics in Quantum Electronics}} \textbf{28}, 1--9 (2022).

\bibitem{pang:2021}
X.~Pang, O.~Ozolins, L.~Zhang, R.~Schatz, A.~Udalcovs, X.~Yu, G.~Jacobsen,
  S.~Popov, J.~Chen, and S.~Lourdudoss, \enquote{Free-space communications
  enabled by quantum cascade lasers,} {\protect\JournalTitle{physica status
  solidi (a)}} \textbf{218}, 2000407 (2021).

\bibitem{pang2022_6gbps}
X.~Pang, R.~Schatz, M.~Joharifar, A.~Udalcovs, V.~Bobrovs, L.~Zhang, X.~Yu,
  Y.-T. Sun, G.~Maisons, M.~Carras, S.~Popov, S.~Lourdudoss, and O.~Ozolins,
  \enquote{Direct modulation and free-space transmissions of up to 6 gbps
  multilevel signals with a 4.65- lt;inline-formula gt; lt;tex-math
  notation="latex" gt;$\mu$ lt;/tex-math gt; lt;/inline-formula gt;m quantum
  cascade laser at room temperature,} {\protect\JournalTitle{Journal of
  Lightwave Technology}} \textbf{40}, 2370--2377 (2022).

\bibitem{Dely:2022}
H.~Dely, T.~Bonazzi, O.~Spitz, E.~Rodriguez, D.~Gacemi, Y.~Todorov, K.~Pantzas,
  G.~Beaudoin, I.~Sagnes, L.~Li, A.~G. Davies, E.~H. Linfield, F.~Grillot,
  A.~Vasanelli, and C.~Sirtori, \enquote{10 gbit s-1 free space data
  transmission at 9~$\mu$m wavelength with unipolar quantum optoelectronics,}
  {\protect\JournalTitle{Laser \& Photonics Reviews}} \textbf{16}, 2100414
  (2022).

\bibitem{sauvage_outdoor_MIR_link_noise}
C.~Sauvage, C.~Robert, B.~Sorrente, F.~Grillot, and D.~Erasme, \enquote{{Study
  of short and mid-infrared telecom links performance for different climatic
  conditions},} in \emph{Environmental Effects on Light Propagation and
  Adaptive Systems II,}  vol. 11153 K.~U. Stein and S.~Gladysz, eds.,
  International Society for Optics and Photonics (SPIE, 2019), pp. 147 -- 155.

\bibitem{note1}
\SI{14}{dB} is the minimum value of attenuation applicable in the case of a
  $P_{out}=\SI{13}{mW}$ to prevent the detector saturation and its degradation.
  \SI{52}{dB} is the maximum value of attenuation applicable in the case of
  $P_{out}=\SI{1.66}{mW}$ to detect the modulation signal.

\bibitem{Gabbrielli:21}
T.~Gabbrielli, F.~Cappelli, N.~Bruno, N.~Corrias, S.~Borri, P.~D. Natale, and
  A.~Zavatta, \enquote{Mid-infrared homodyne balanced detector for quantum
  light characterization,} {\protect\JournalTitle{Opt. Express}} \textbf{29},
  14536--14547 (2021).

\bibitem{Rajagopal:12}
S.~Rajagopal, R.~D. Roberts, and S.-K. Lim, \enquote{{IEEE} 802.15.7 visible
  light communication: modulation schemes and dimming support,}
  {\protect\JournalTitle{IEEE Communications Magazine}} \textbf{50}, 72--82
  (2012).

\bibitem{Seminara2020}
M.~{Seminara}, T.~{Nawaz}, S.~{Caputo}, L.~{Mucchi}, and J.~{Catani},
  \enquote{{Characterization of Field of View in Visible Light Communication
  Systems for Intelligent Transportation Systems},} {\protect\JournalTitle{IEEE
  Photonics Journal}} \textbf{12}, 3005620 (2020).

\bibitem{ASHOK:2019}
P.~Ashok and M.~{Ganesh Madhan}, \enquote{Performance analysis of various pulse
  modulation schemes for a fso link employing gain switched quantum cascade
  lasers,} {\protect\JournalTitle{Optics \& Laser Technology}} \textbf{111},
  358--371 (2019).

\bibitem{Malik:15}
A.~Malik and P.~Singh, \enquote{Free space optics: Current applications and
  future challenges,} {\protect\JournalTitle{International Journal of Optics}}
  (2015).

\bibitem{2019iv}
A.~K. Majumdar, \emph{Optical Wireless Communications for Broadband Global
  Internet Connectivity} (Elsevier, 2019).

\bibitem{Meucci21}
M.~Meucci, M.~Seminara, T.~Nawaz, S.~Caputo, L.~Mucchi, and J.~Catani,
  \enquote{Bidirectional vehicle-to-vehicle communication system based on vlc:
  Outdoor tests and performance analysis,} {\protect\JournalTitle{IEEE
  Transactions on Intelligent Transportation Systems}} pp. 1--11 (2021).

\bibitem{Khalili05}
R.~{Khalili} and K.~{Salamatian}, \enquote{A new analytic approach to
  evaluation of packet error rate in wireless networks,} in \emph{3rd Annual
  Communication Networks and Services Research Conference (CNSR'05),}  (2005),
  pp. 333--338.

\bibitem{wilson:96}
S.~Wilson, \emph{Digital Modulation and Coding} (Prentice Hall, 1996).

\bibitem{book:kaushal2017}
H.~Kaushal, V.~Jain, and S.~Kar, \emph{Free space optical communication}
  (Springer, 2017).

\bibitem{ITU1}
\enquote{Propagation data required for the design of terrestrial free-space
  optical links,} {\protect\JournalTitle{Recommendation {ITU-R}}}  (2012).

\bibitem{ITU2}
\enquote{Prediction methods required for the design of terrestrial free-space
  optical links,} {\protect\JournalTitle{Recommendation {ITU-R}}}  (2012).

\bibitem{ricklin:2006}
J.~C. Ricklin, S.~M. Hammel, F.~D. Eaton, and S.~L. Lachinova,
  \enquote{Atmospheric channel effects on free-space laser communication,}
  {\protect\JournalTitle{Journal of Optical and Fiber Communications Reports}}
  \textbf{3}, 111--158 (2006).

\bibitem{henniger:2010}
H.~Henniger and O.~Wilfert, \enquote{An introduction to free-space optical
  communications.} {\protect\JournalTitle{Radioengineering}} \textbf{19}
  (2010).

\bibitem{Zhao:2019}
B.-B. Zhao, X.-G. Wang, J.~Zhang, and C.~Wang, \enquote{Relative intensity
  noise of a mid-infrared quantum cascade laser: insensitivity to optical
  feedback,} {\protect\JournalTitle{Opt. Express}} \textbf{27}, 26639--26647
  (2019).

\bibitem{borri2011frequency}
S.~Borri, S.~Bartalini, P.~C. Pastor, I.~Galli, G.~Giusfredi, D.~Mazzotti,
  M.~Yamanishi, and P.~De~Natale, \enquote{Frequency-noise dynamics of
  mid-infrared quantum cascade lasers,} {\protect\JournalTitle{IEEE Journal of
  Quantum Electronics}} \textbf{47}, 984--988 (2011).

\bibitem{bartalini2011}
S.~Bartalini, S.~Borri, I.~Galli, G.~Giusfredi, D.~Mazzotti, T.~Edamura,
  N.~Akikusa, M.~Yamanishi, and P.~De~Natale, \enquote{Measuring frequency
  noise and intrinsic linewidth of a room-temperature dfb quantum cascade
  laser,} {\protect\JournalTitle{Optics express}} \textbf{19}, 17996--18003
  (2011).

\bibitem{Stern:04}
H.~Stern, S.~Mahmoud, and L.~Stern, \emph{Communication Systems: Analysis and
  Design} (Pearson Prentice Hall, 2004).

\bibitem{hitrandat}
{Harvard--Smithsonian Center for Astrophysics (CfA)}, \emph{The {HITRAN}
  database} (2013).

\bibitem{hotw}
{Harvard--Smithsonian Center for Astrophysics (CfA), V.~E.~Zuev Insitute of
  Atmosperic Optics (IAO)}, \emph{HITRAN on the Web} (2019).

\bibitem{Rothman:2013}
L.~Rothman, I.~Gordon, Y.~Babikov, A.~Barbe, D.~{Chris Benner}, P.~Bernath,
  M.~Birk, L.~Bizzocchi, V.~Boudon, L.~Brown, A.~Campargue, K.~Chance,
  E.~Cohen, L.~Coudert, V.~Devi, B.~Drouin, A.~Fayt, J.-M. Flaud, R.~Gamache,
  J.~Harrison, J.-M. Hartmann, C.~Hill, J.~Hodges, D.~Jacquemart, A.~Jolly,
  J.~Lamouroux, R.~{Le Roy}, G.~Li, D.~Long, O.~Lyulin, C.~Mackie, S.~Massie,
  S.~Mikhailenko, H.~Müller, O.~Naumenko, A.~Nikitin, J.~Orphal, V.~Perevalov,
  A.~Perrin, E.~Polovtseva, C.~Richard, M.~Smith, E.~Starikova, K.~Sung,
  S.~Tashkun, J.~Tennyson, G.~Toon, V.~Tyuterev, and G.~Wagner, \enquote{The
  hitran2012 molecular spectroscopic database,} {\protect\JournalTitle{Journal
  of Quantitative Spectroscopy and Radiative Transfer}} \textbf{130}, 4--50
  (2013). HITRAN2012 special issue.

\bibitem{kneizys1980atmospheric}
F.~X. Kneizys, \emph{Atmospheric Transmittance/radiance, Computer Code LOWTRAN
  5}, 697 (Optical Physics Division, Air Force Geophysics Laboratory, 1980).

\bibitem{MALIK:2020}
S.~Malik and P.~K. Sahu, \enquote{Free space optics/millimeter-wave based
  vertical and horizontal terrestrial backhaul network for 5g,}
  {\protect\JournalTitle{Optics Communications}} \textbf{459}, 125010 (2020).

\bibitem{HANDURA2016}
M.~Handura, K.~Ndjavera, C.~Nyirenda, and T.~Olwal, \enquote{Determining the
  feasibility of free space optical communication in namibia,}
  {\protect\JournalTitle{Optics Communications}} \textbf{366}, 425--430 (2016).

\bibitem{valley1980isoplanatic}
G.~C. Valley, \enquote{Isoplanatic degradation of tilt correction and
  short-term imaging systems,} {\protect\JournalTitle{Applied Optics}}
  \textbf{19}, 574--577 (1980).

\end{thebibliography}

\end{document}